\documentstyle[11pt,newpasp,twoside,epsf]{article}
\newcommand{\mss}{mag arcsec$^{-2}$}

\newcommand{\plm}{$\pm$ }

\newcommand{\muo}{\rm ${\mu}$(0) }

\newcommand{\etal}{{\it et.al.}\ }

\newcommand{\MLsol}{${M_{\odot}}/{L_{\odot}}$\ }
\newcommand{\Msol}{$M_{\odot}$\ }
\newcommand{\Zsol}{$Z_{\odot}$\ }

\input{epsf}
\input{rotate}
\begin{document}
\markboth{K. O'Neil}{The Morphology, Color, and Gas Content of Low Surface Brightness Galaxies}

\title{The Morphology, Color, and Gas Content of\\
Low Surface Brightness Galaxies}
\author{K. O'Neil}
\affil{Arecibo Observatory HC3 Box 53995 Arecibo, PR 00612\\ koneil@naic.edu}

\begin{abstract}
Recent surveys have discovered hundreds of low surface brightness galaxies,
systems with central surface brightness fainter than 22.0 B mag arcsec$^{-2}$,
in the local universe.  Plots of the surface brightness distribution --
that is, the space density of galaxies plotted against central surface
brightness -- show a flat space density distribution from the canonical
Freeman value of 21.65 through the current observational limit of
25.0 B mag arcsec$^{-2}$.  It is therefore extremely important
to understand these diffuse systems if we wish to understand galaxy
formation and evolution as a whole.  This talk is a review of 
both the known properties of low surface brightness galaxies and of 
popular theories describing the formation and evolution of these enigmatic systems.
\end{abstract}

\keywords{galaxies:morphology -- galaxies:evolution -- galaxies:formation --
galaxies:low surface brightness -- galaxies:surveys -- galaxies:stars}

\section{Introduction}

\begin{figure}[ht]
\centerline{
\sbox{1}{
\epsfxsize=2.3in
\epsffile{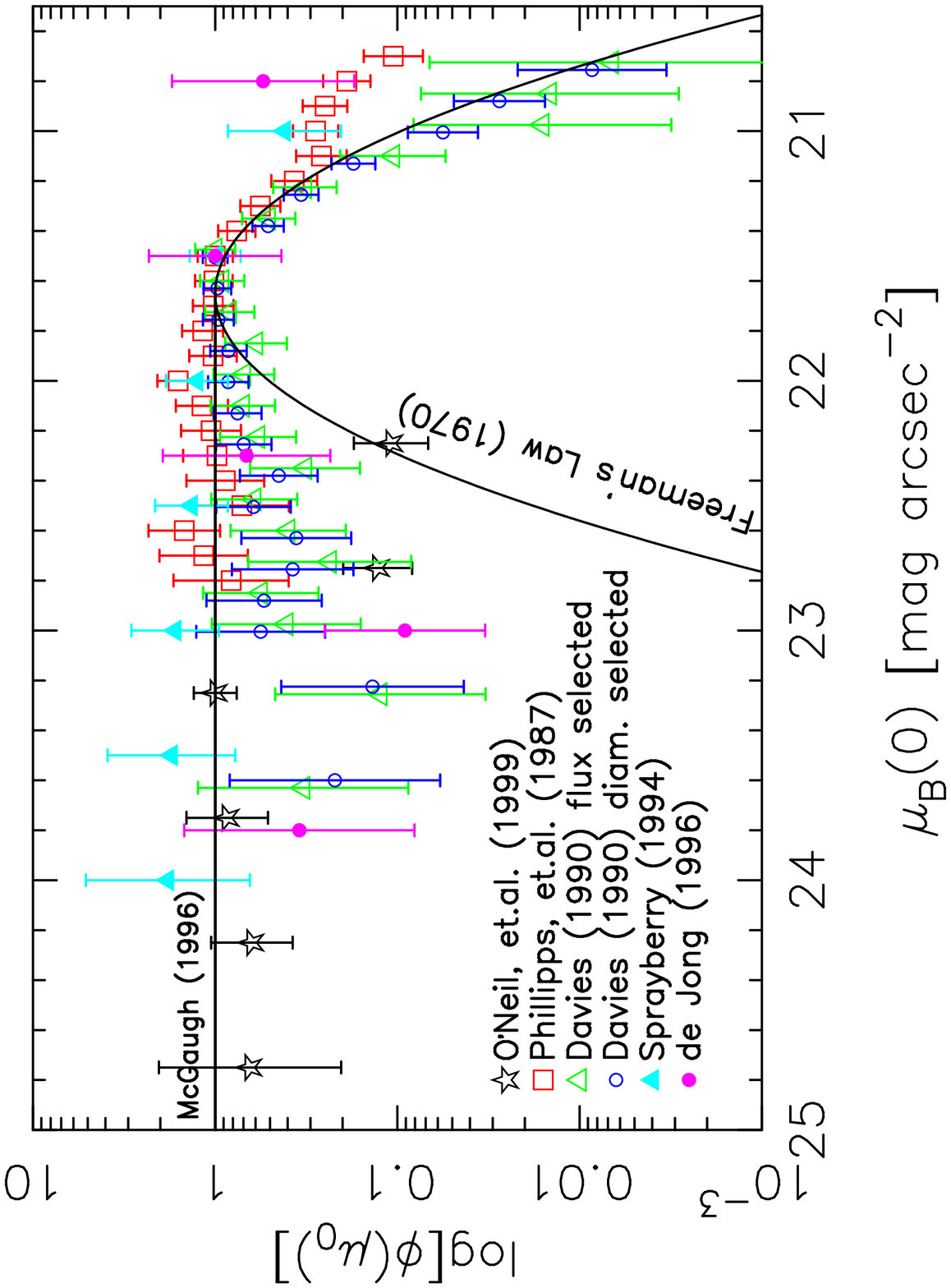}}
\rotr{1}
\sbox{1}{
\epsfxsize=2.3in
\epsffile{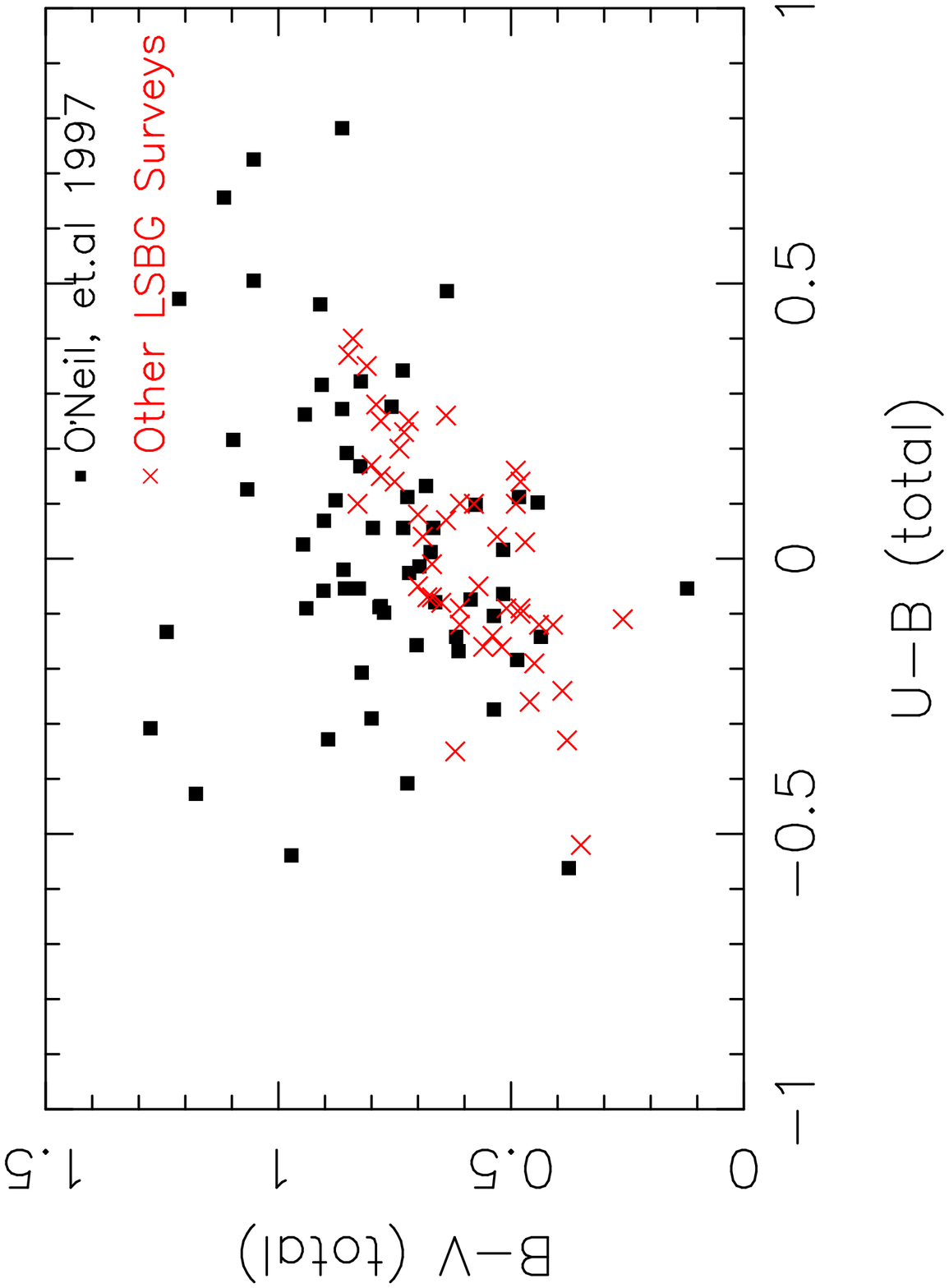}}
\rotr{1}
}
\vskip 0.1in
\caption{(a) The number density of galaxies in the local universe, with $\phi$ normalized to one
(O'Neil \& Bothun 2000).  (b) A representative sample of LSB galaxy colors, from O'Neil, 
\etal\ (1997b).\label{fig:sbdist}}\vskip -0.1in
\end{figure}

The importance of low surface brightness (LSB) galaxies in the local universe has
recently been emphasized in a study by O'Neil \& Bothun (2000) which has extended
the known distribution of galaxies in the local universe.
Their result is a flat
surface brightness distribution function from the Freeman value of
21.65 \plm 0.30 to the survey limit of 25.0 B \mss, more than
10$\sigma$ away (Figure~\ref{fig:sbdist}a).  This indicates that the majority of
the galaxies, and potentially the majority of the baryons in the local universe,
are contained in gravitational potentials only dimly lit by the embedded galaxy.
Realizing that most galaxies are optically diffuse, then, it becomes extremely
important to understand these systems if we wish to understand galaxy 
formation and evolution as a whole.  With this in mind I wish to undertake
a brief review both of our current understanding of LSB galaxy properties
as well as a review of some popular ideas behind the formation of these 
enigmatic systems.

\section{What We (think) We Know about LSB Galaxies}

{\bf LSB galaxy colors:}  Contrary to what was first believed, the colors
of LSB galaxies range across the entire high surface brightness (HSB) galaxy 
spectrum, including what may be the bluest galaxy known (UGC 12695, with 
$U - I\:=\:-0.2$) as well has some fairly red systems
($B - V\:>\:1.0$) (Figure~\ref{fig:sbdist}b) (O'Neil, \etal
1997a, 1997b, 1998). 
Although currently it appears that LSB galaxy colors do
not quite extend into the extremely
red colors found in some HSB galaxies, this is likely due more to small
number statistics rather than to an actual lack of extremely red LSB systems.

\begin{figure}
\centerline{
\hskip 0.7in
\sbox{1}{
\epsfxsize=2.3in
\epsffile{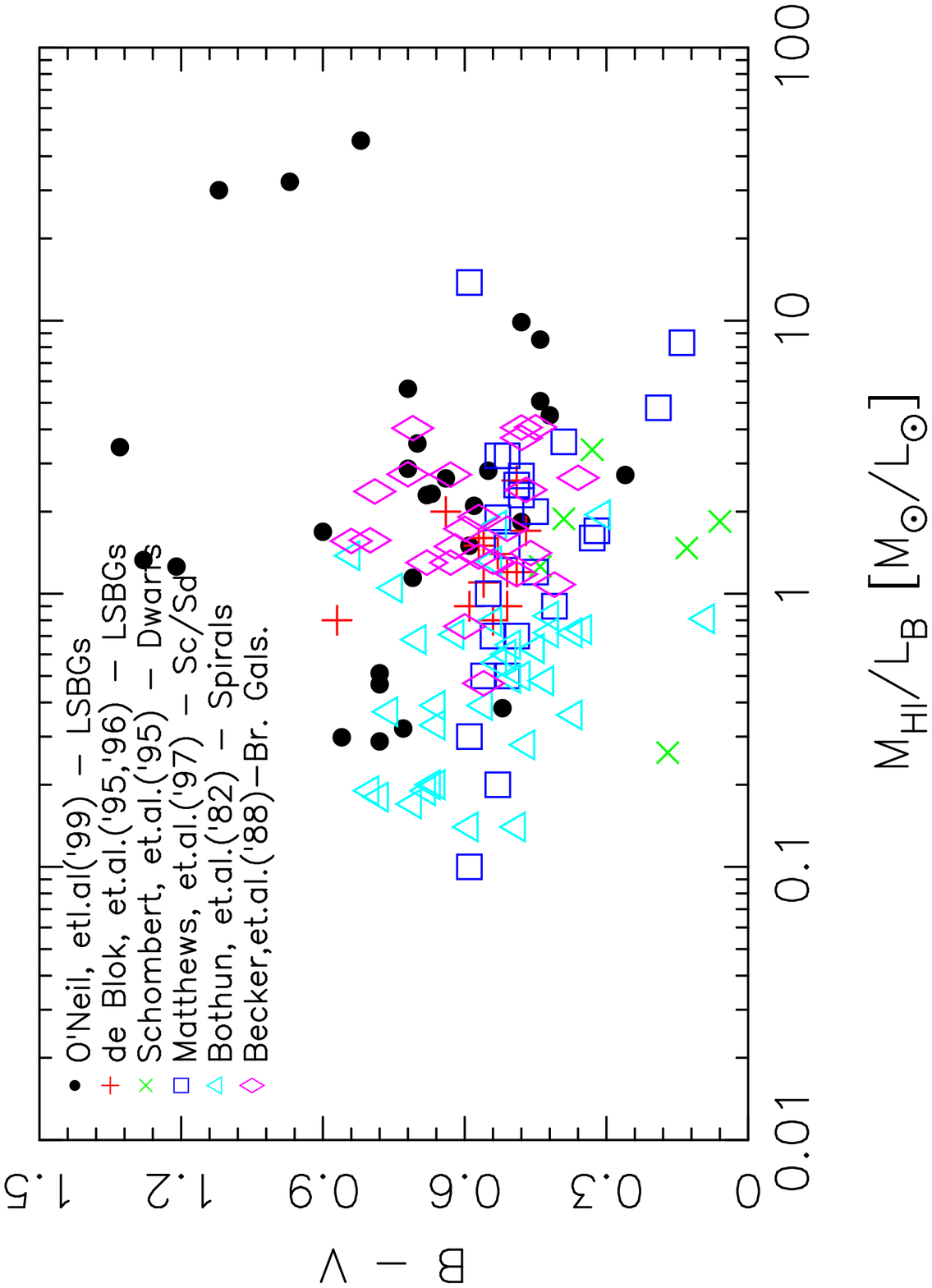}}
\rotr{1}
\sbox{1}{
\epsfxsize=2.3in
\epsffile{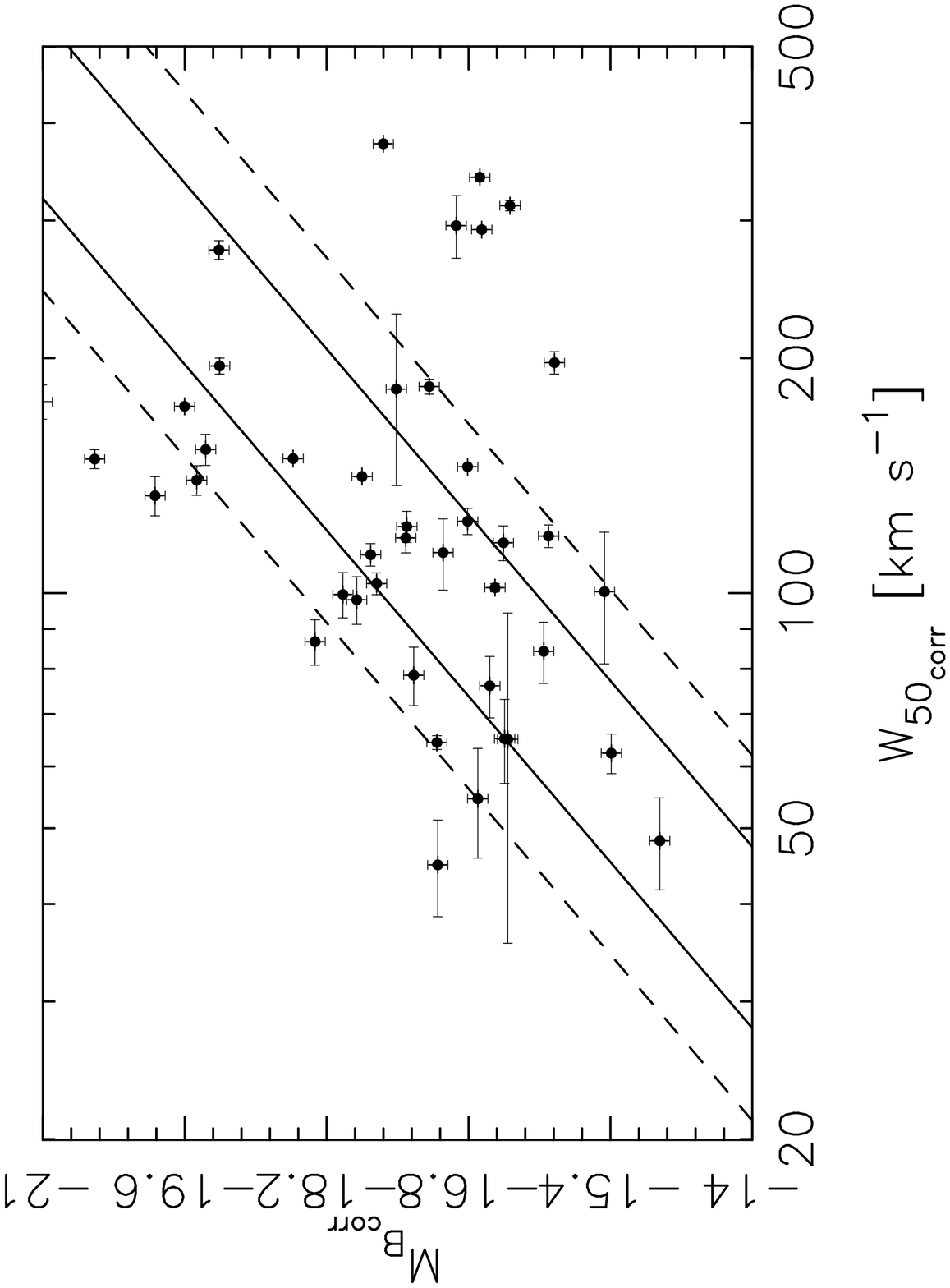}}
\rotr{1}}
\vskip 0.1in
\caption{(a) Color versus mass-to-luminosity ratio for a variety of galaxy types (from
OBS).  (b) Galaxies from the sample of OBS.
The solid and dashed lines are the 1$\sigma$ and 2$\sigma$ fits
to the LSB galaxy T-F relation of Zwaan, \etal\ (1995). \label{fig:gascol}}\vskip -0.1in
\end{figure}

{\bf Gas-to-luminosity ratio of LSB galaxies:}  The gas-to-luminosity ratio
(M$_{HI}$/L$_B$) of LSB galaxies spans an extremely large range, from fairly low
(M$_{HI}$/L$_B$ = 0.1 \MLsol) through what may be the highest M$_{HI}$/L$_B$
galaxies known ([OBC97] N9-2, M$_{HI}$/L$_B$ = 46 \MLsol).
Additionally, if any trend can be seen between
the galaxies' color and gas content it is that there may be an {\it increase} in
their M$_{HI}$/L$_B$ with redder color (Figure~\ref{fig:gascol}a)
(O'Neil, Bothun, \& Schombert 2000, OBS from now on).

{\bf Tully-Fisher relation:}  Although previous studies have shown LSB galaxies
to follow a slightly broadened version of the standard Tully-Fisher (T-F)
relation defined by HSB galaxies (Zwaan, \etal 1995), a recent study of over 40
LSB galaxies found no significant correlation between LSB galaxy
velocity widths and absolute magnitudes, with
only 40\% of the sample falling within 1$\sigma$ of the previously defined
LSB T-F relation (Figure~\ref{fig:gascol}b).  At the least, then,
there is a significant population of LSB galaxies which do not adhere to the 
T-F relation (OBS).
 
{\bf Rotation curves:}  The rotation curves of LSB galaxies 
have been shown to rise more slowly than similar HSB galaxies
Using `standard' values for the stellar mass-to-luminosity
ratio, as taken from HSB galaxies ($\Upsilon_*$ = 1 -- 3), this leads to the conclusion
that many LSB galaxies have a baryonic mass fraction up to 3$\times$ less than
HSB galaxies with the same velocity width (i.e. Swaters, \etal\ 2000;
Van Zee, \etal\ 1998; de Blok \& McGaugh 1997). 

\section{What Are LSB Galaxies?}

{\bf The faded version of HSB galaxies?} No.  LSB galaxies often have both very blue colors 
and very low metallicities ($B-V$ $<$ 0.2, Z $<$ 0.01\Zsol), precluding the possibility 
that LSB galaxies are primarily composed of an old stellar population.  As a caveat,
though, it should be noted that a number of very red LSB galaxies have now been
found, and these could be faded HSB galaxies.  If this is
correct, though, all other LSB galaxies (i.e. those which do not have extremely red colors)
would have to be explained, as well as why there are two separate populations of
LSB galaxies (O'Neil, \etal\ 1997a, 1997b).

{\bf ``Stretched out'' HSB galaxies?} No.  Current theories describing LSB
galaxies as extending further into their dark matter haloes than similar HSB
galaxies predict that LSB galaxies will follow either a universal T-F
relation or one which is unique at each \muo.
These theories therefore cannot account for the galaxies of OBS
which fall well off the T-F relation, with no correlation between
\muo\ and residual error.  Additional problems with the models can 
(depending on which models are considered) include:
difficulty matching the observed shape of LSB galaxy rotation curves;
inability to allow for high gas fraction, red
LSB galaxies, (i.e. Dalcanton, \etal\ 1997; McGaugh \& de Blok 1998;
Avila-Reese \& Firmani 2000; McGaugh 1999).

{\bf A completely new type of galaxy?}  No.  Although this idea could justify 
ignoring LSB galaxies when determining theories of galaxy formation and evolution,
no evidence has been seen for LSB galaxies to be anything but a continuation
of the HSB galaxy spectrum.  There is a smooth transition between LSB and HSB galaxies
in surface brightness
and complete overlap in LSB and HSB galaxy colors, scale lengths, mass, 
luminosity, etc. (i.e. Bell \& de Blok 2000; O'Neil, \etal\ 1997a, 1997b).

{\bf Galaxies with a different stellar population?}  Maybe.  Although this
is not a popular idea, as having an IMF which depends on galaxy
properties (i.e. surface density) adds complication to models of
galaxy evolution, this theory has not yet been disproved.  
The gas density of LSB galaxies is typically at or below the nominal 
threshold for star formation, as set by the Toomre criterion (i.e. Van Zee, \etal\ 1998;
de Blok, McGaugh, \& Van der Hulst 1996).
With this in mind, it would be suprising if LSB galaxies' IMF was not at least
somewhat affected by their low density.
Additionally, recent HST WFPC-2 studies of three nearby LSB dE galaxies failed to find
evidence for a significant number of red giants ($<$ 13 per 10 pc$^2$, as opposed
to the 100s per 10 pc$^2$ typically found in HSB galaxies (O'Neil, \etal\ 1999)).
Using these two ideas -- the low gas density and the lack of evidence for significant
numbers of giant branch stars -- we can construct a toy model wherein no stars
greater than 2\Msol are allowed to form.  When this is done, not only are LSB
galaxy colors, gas fractions, etc. readily matched, but it is also remarkably
easy to form both red and blue galaxies which do not follow the canonical 
T-F relation (see OBS).  Additionally,
the addition of a large number of small stars to any galaxy dramatically increases
the galaxy's stellar mass-to-luminosity ratio ($\Upsilon_*$) and can dramatically
decrease the total amount of dark matter needed in LSB systems (Swaters, \etal\ 2000;
OBS).  Although 
these models are admittedly extremely oversimplified, they pave the way for
further studies into this idea, and currently appear to be the best theory going.

\end{document}